\documentclass[prl,a4paper,superscriptaddress,twocolumn,showpacs,amsmath,amssymb,floatfix,amstex]{revtex4-1}%

\usepackage[T1]{fontenc}
\usepackage[utf8]{inputenc} 
\usepackage[english]{babel}    

\usepackage{graphicx}

\usepackage{hyperref}
\hypersetup{colorlinks=true,allcolors=blue}

\usepackage{xcolor}
\usepackage{physics}

\usepackage{grffile}
\usepackage{dcolumn}
\usepackage{mathptmx}
\usepackage{ulem}

\newcommand{\e}[1]{\mathrm{e}^{#1}}

\begin{document}
\title{Coherent Forward Scattering Peak and Multifractality}
\author{Maxime Martinez}
\affiliation{Laboratoire de Physique Th\'eorique, Universit\'e de Toulouse, CNRS, UPS, France}
\author{Gabriel Lemari\'e}
\affiliation{Laboratoire de Physique Th\'eorique, Universit\'e de Toulouse, CNRS, UPS, France}
\affiliation{MajuLab, CNRS-UCA-SU-NUS-NTU International Joint Research Unit, Singapore}
\affiliation{Centre for Quantum Technologies, National University of Singapore, Singapore}
\author{Bertrand Georgeot}
\affiliation{Laboratoire de Physique Th\'eorique, Universit\'e de Toulouse, CNRS, UPS, France}
\author{Christian Miniatura}
\affiliation{MajuLab, CNRS-UCA-SU-NUS-NTU International Joint Research Unit,Singapore}
\affiliation{Centre for Quantum Technologies, National University of Singapore, Singapore}
\affiliation{Universit\'e  C\^ote  d'Azur,   CNRS,   INPHYNI,   Nice,   France}
\affiliation{Department of Physics, National University of Singapore, Singapore}
\affiliation{School  of  Physical  and  Mathematical  Sciences,  Nanyang  Technological University, Singapore}
\author{Olivier Giraud}
\affiliation{Université Paris-Saclay, CNRS, LPTMS, 91405 Orsay, France}

\date{\today}

\begin{abstract}
It has recently been shown that interference effects in disordered systems give rise to two non-trivial structures: the coherent backscattering (CBS) peak, a well-known signature of interference effects in the presence of disorder, and the coherent forward scattering (CFS) peak, which emerges when Anderson localization sets in.  We study here the CFS effect in the presence of quantum multifractality, a fundamental property of several systems, such as the Anderson model at the metal-insulator transition.  We focus on Floquet systems, and find that the CFS peak shape and its peak height dynamics are generically controlled by the multifractal dimensions $D_1$ and $D_2$, and by the spectral form factor. We check our results 
using a 1D Floquet system whose states have multifractal properties controlled by a single parameter. Our predictions are fully confirmed by numerical simulations and analytic perturbation expansions on this model. Our results, which we believe to be generic, provide an original and direct way to detect and characterize multifractality in experimental systems.
\end{abstract}

%
\maketitle

{\it Introduction.}
In the field of quantum transport, the coherent backscattering (CBS) effect is a well-known signature of interference effects that emerges at a time of the order of the elastic scattering time, and survives configuration average in time-reversal symmetric disordered systems \cite{AkMon,Cher12,Jen12,Lab12}.  It is visible as a peak in momentum space for spatially disordered systems and in position space for the type of Floquet systems that we consider in the present Letter~\cite{Lema17}. Recently, it was discovered that, in the presence of Anderson localization, CBS was further accompanied by the emergence of a coherent forward scattering (CFS) peak (which actually arises even without time-reversal symmetry), leading to a twin-peak structure breaking ergodicity in the long-time limit \cite{Kar12}. 

The CFS peak is in fact a smoking gun of strong localization \cite{Kar12, Lee14, Ghosh14, Mick14, Ghosh15, Ghosh17, Lema17, Marin18}.
In particular, it was shown \cite{Ghosh17} that it could be used to monitor the metal-insulator Anderson transition, as it vanishes in the metallic phase and is fully developed in the localized regime.
It was even suggested  \cite{Ghosh17} that at the transition, where there exists non-ergodic delocalized
states with multifractal properties \cite{EveMir08}  (i.e.~scale-invariant fluctuations characterized by a continuous set of fractal dimensions $D_q$) the CFS peak might embody these multifractal properties. 
In this Letter, we consider the particular case of Floquet systems where localization and/or multifractality occur in momentum space. This type of dynamical systems is extremely convenient for extensive numerical studies and has been already implemented in several experiments, see e.g. ~\cite{Moore94, Chabe2008}. For these kicked systems, we demonstrate that the height and shape of the CFS peak (a dynamical observable in position space) give a direct and remarkable access to the multifractal dimensions $D_1$ and $D_2$ (a static property of the Floquet eigenstates in momentum space). 
Our general predictions are very well corroborated by numerical simulations and analytical perturbative expansions on the Ruijsenaars-Schneider model \cite{RS86}, a dynamical system where all states have multifractal properties controlled by a single parameter. Since our results are based on a very general theoretical framework and derived using well-supported arguments, we believe that they should apply to any critical disordered system.

This work paves the way to a first robust experimental study of quantum multifractality, that remains very hard to observe despite a huge theoretical interest (see the pioneering experiments \cite{Morgen03, RicRou10} in a quantum setting and \cite{Faez09} in a classical setting). 
Indeed, the CFS peak is a direct experimental observable that has recently been observed with cold atom experiments in the localized regime \cite{Hai18}.


\begin{figure}
\begin{center}
\includegraphics[width=0.99\linewidth]{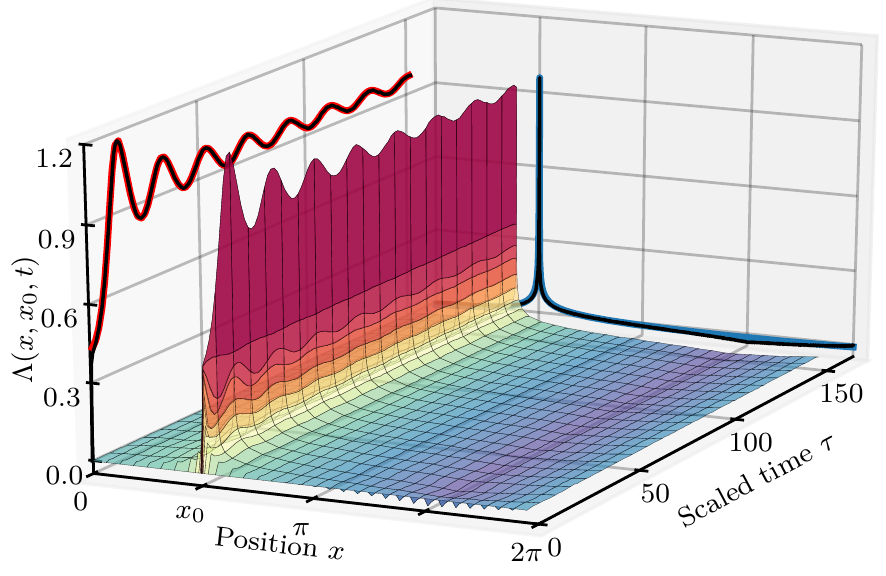}
\caption{Contrast of the CFS peak, Eq.~\eqref{deflambda}, with scaled time $\tau=2\pi t/N$, for the RS model with $N=4096$ and $a=0.322$. Left projection: height of the CFS peak. The theoretical prediction (red solid line), see text below Eq.~\eqref{lambdatempx0}, matches perfectly the numerical data (black solid line). Right projection: shape of the CFS peak at $\tau =150$. The black solid line represents the numerical data and the blue solid line is Eq.~\eqref{lambdainfx} rescaled (see text).
\label{fig3D}}
\end{center}
\end{figure}




{\it The Ruijsenaars-Schneider (RS) model.}
The RS model \cite{RS86} is a variant of the kicked rotor~\cite{Chir79, Izra90}, a paradigmatic model of quantum chaos which exhibits Anderson localization in momentum space.
It is a 1D Floquet system whose corresponding Hamiltonian reads $H=p^2/2 - 2\pi a \, [x (\textrm{mod}\, 2\pi)] \sum_{n=-\infty}^{\infty} \delta(t-n)$, featuring a periodically-kicked sawtooth potential with strength $2\pi a$.
The difference with the kicked rotor comes from the spatial discontinuities of the sawtooth potential, inducing long-range hopping between momentum basis states, which breaks standard exponential localization \cite{SelVerZir85,Mir96,EveMir08,DenKra18, NosKha19}. Saliently, the RS model displays  multifractal eigenstates~\cite{OG006,OG032, OG034,Garcia05}.
The dynamics of such periodically-kicked systems is captured by the Floquet operator $U$ over one period, whose eigenvectors $\ket{\varphi_\alpha}$ are associated with quasi-energies $\omega_\alpha \in [-\pi, \pi[$, so that $U^t \ket{\varphi_\alpha} = \e{i \omega_\alpha t} \ket{\varphi_\alpha}$. 
For the RS model, $U$ can be written explicitly in the momentum basis $| p \rangle$ (with integer $p$, $0\leq p \leq N-1$) as
\begin{equation}
\label{ruij}
\langle p| U | p' \rangle = U_{pp'}= \frac{e^{i \phi_p}}{N}\frac{1-e^{2\pi i a}}{1-e^{2\pi i(p'-p+a)/N}},
\end{equation}
where the dynamical phases $\phi_p$ can be taken as randomly distributed over $[0,2\pi[$. In the following, we will denote disorder average by bracketed terms $\langle (\cdot\cdot\cdot) \rangle$. This random matrix ensemble has been intensively studied in many branches of theoretical physics and mathematics  \cite{OG022,OG030,OG014,OG042,BraSas97,RS86,BogSch04,BogDubSch09,refm1,refm2}. In particular, it breaks time-reversal symmetry, so that the usual CBS effect is destroyed \cite{Lema17,Hai18}, and its spectrum displays intermediate statistics \cite{OG030}. 

It is important to recall that the eigenstates of RS are multifractal in \textit{momentum} space \cite{OG032, OG034}. This is characterized by the anomalous scaling of the disorder-averaged moments of the wavefunctions, $\langle\sum_p|\varphi_\alpha(p)|^{2q}\rangle\sim N^{-D_q(q-1)}$, where $\varphi_\alpha(p)\equiv \braket{p}{\varphi_\alpha}$ and $D_q$ are the multifractal dimensions. 
In the RS model, the parameter $a$ controls the nature of the eigenstates \cite{OG032}: when $a$ goes from $0$ to $1$ the system goes from the regime of strong multifractality ($D_q \ll 1$) to the regime of weak multifractality ($D_q \sim 1$), making it an ideal testbed for our theory.\\


\begin{figure*}[!t]
\centering
\includegraphics{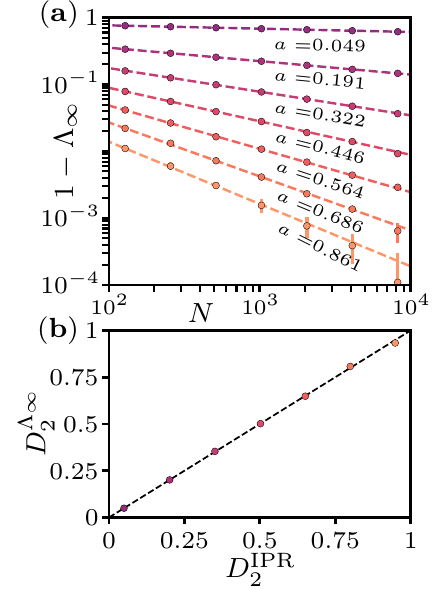}
\includegraphics[scale=1]{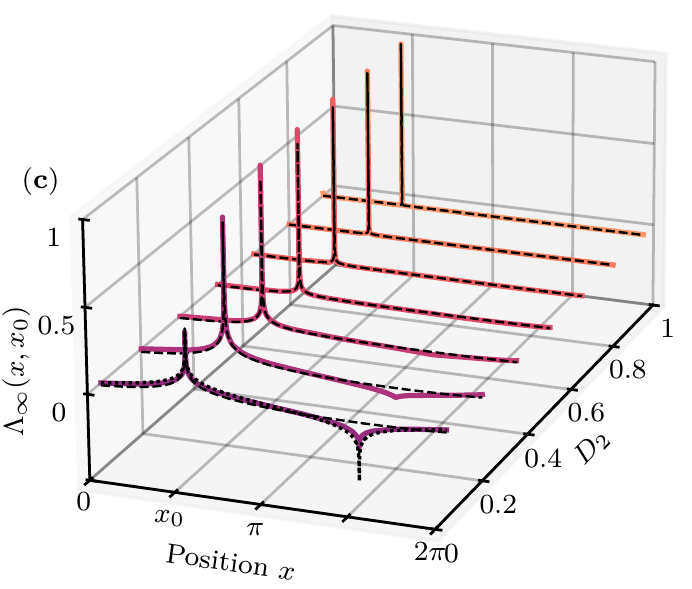}
\includegraphics[scale=1]{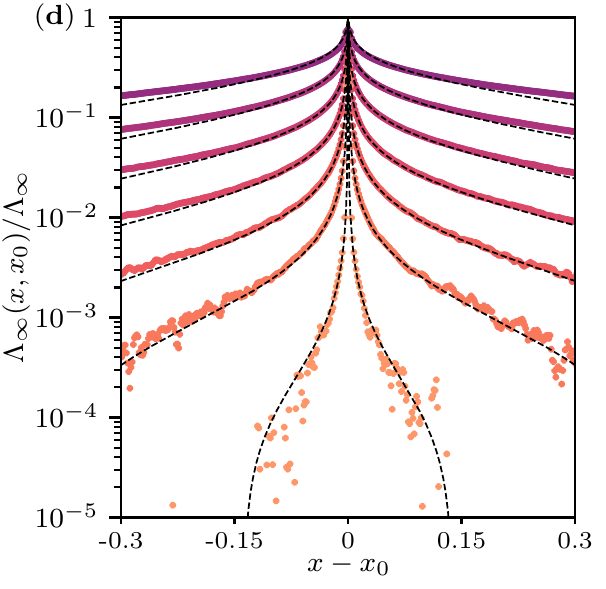}
\caption{Contrast of the CFS peak at infinite time. (a) Finite-size scaling of $\Lambda_\infty$ for different values of $a$.  The dashed lines correspond to the nonlinear fit $1-\Lambda_\infty=\alpha N^{-D_2^{\Lambda_\infty}}$ with parameters $\alpha$ and $D_2^{\Lambda_\infty}$. (b) Comparison between the values of $D_2$ extracted from a finite-size scaling of $\Lambda_\infty$ in (a) and from the inverse participation ratio (IPR) defined by the right-hand side of Eq.~\eqref{lambdainfx0}. The two coincide very well. When not visible, error bars are smaller than the symbol size. (c) CFS peak contrast for the values of $a$ used in (a), with $N=4096$. Solid lines are numerical data, black dashed lines are Eq.~\eqref{lambdainfx} rescaled to minimize the difference with numerics on a spatial range $\Delta x =0.5$ around $x=x_0$ (excluding the point at $x_0$), black dotted line is Eq.~\eqref{finaloverlap}. (d) Zoom-in of (c) around $x_0$ and smoothed over $\delta x =0.007$.
\label{fig:contraste-infini}}
\end{figure*}

{\it The CFS contrast.} 
The RS model above is only an example of a more general class of systems which can be described by an evolution operator $U$ with Floquet states localized or multifractal in momentum space. For such systems, the CFS interference phenomenon takes place in position space \cite{Lema17}. 
We introduce the position basis $|x \rangle$ ($x=2\pi n/N$, $0\leq n \leq N-1$) which is related to the momentum basis $|p \rangle$ defined above by Fourier transform $\langle x|p\rangle=\exp(ipx)/\sqrt{N}$.

In the following, we thus consider the time evolution of the system starting from some initial state $|x_0 \rangle$ in position space and analyze the disorder-averaged position distribution after $t$ iterations of the map $U$, namely $\langle\vert\bra{x}U^t\ket{x_0}\vert^2\rangle$. After an initial transient regime, it features a peak around the initial value $x=x_0$, the CFS peak. 
To single out this interference effect resisting disorder average, we introduce the \textit{contrast} $\Lambda(x,x_0,t)$ as the relative difference between the quantum probability distribution $\langle\vert\bra{x}U^t\ket{x_0}\vert^2$ and the classical, interference-free, long-time limit $1/N$:
\begin{equation}
\label{deflambda}
\Lambda (x,x_0,t)=\frac{\langle\vert\bra{x}U^t\ket{x_0}\vert^2\rangle-1/N }{1/N}.
\end{equation}
The time behavior of the contrast is illustrated in Fig.~\ref{fig3D} in the case of the RS model. A peak emerges at short times around $x=x_0$, its height oscillates (left projection in Fig.~\ref{fig3D}) and eventually stabilizes (right projection).

Expanding over eigenstates of $U$, the contrast \eqref{deflambda} writes
\begin{equation}
\label{defdensity}
\Lambda (x,x_0,t) = N\sum_{\alpha\beta}\Big\langle e^{i \omega_{\alpha\beta} t}\varphi^*_\alpha(x)\varphi_\alpha(x_0)\varphi^*_\beta(x_0)\varphi_\beta(x)\Big\rangle-1,
\end{equation}
where $\varphi_\alpha(x)\equiv \braket{x}{\varphi_\alpha}$ and $\omega_{\alpha\beta}=[\omega_\alpha-\omega_\beta] (\textrm{mod} \, 2\pi) \in [-\pi,\pi[ $. Note that in Eq.~\eqref{defdensity}, $t$ can be considered a continuous variable: In the following we shall therefore resort to the usual Fourier transform rather than the discrete one. 

At long times, only the diagonal part $\alpha=\beta$ in Eq.~\eqref{defdensity} survives, giving, for fixed system size $N$, the stationary limit
\begin{equation}
\label{lambdaxx0inf}
\Lambda_\infty(x,x_0) = N \sum_\alpha\big\langle\qty|\varphi_\alpha(x)|^2\qty|\varphi_\alpha(x_0)|^2 \big\rangle-1.
\end{equation}
The time dependence of $\Lambda (x,x_0,t) $ is fully encapsulated in the off-diagonal terms $\alpha \neq \beta$.
The function $F(x,x_0,t)=\Lambda (x,x_0,t)-\Lambda_\infty(x,x_0)$ that governs the time dynamics of the contrast is given by the inverse Fourier transform of
\begin{equation}
\label{defb}
\hat{F}(x,x_0,\omega)=2\pi N\!\!\! \displaystyle{\sum_{\alpha\neq\beta}}\!\!\!\big\langle\delta(\omega-\omega_{\alpha\beta})\varphi^*_\alpha(x)\varphi_\alpha(x_0)\varphi^*_\beta(x_0)\varphi_\beta(x)\big\rangle.
\end{equation}
In what follows, we will first analyze the stationary (i.e. $t\rightarrow \infty$) contrast $\Lambda_\infty(x,x_0)$ for finite $N$, discuss its peak value at $x=x_0$ and its shape around $x_0$. Then, we will discuss the time dynamics of the peak at $x=x_0$, given by $F(x_0,x_0,t)$ and show that the limits of large times $t$ and large system sizes $N$ do not commute. These stationary distribution and time dynamics are illustrated in Fig.~\ref{fig3D} for the RS model. 

At this point, our strategy is to connect these dynamical quantities expressed in $x$-space to the known multifractal properties of the Floquet eigenstates in $p$-space. For this, we use the spatial Fourier transform and introduce the 4-point correlator in momentum space:
\begin{equation}
\label{lambdaxx0infexp}
C_{\alpha\beta} (p_1,p'_1,p_2,p'_2)  =\big\langle \varphi_\alpha(p_1) \varphi_\alpha^*(p'_1) \varphi_\beta(p_2) \varphi_\beta^*(p'_2)\big\rangle.
\end{equation}
As is well-known, the scaling properties of the correlator \eqref{lambdaxx0infexp} encapsulate the multifractal dimensions \cite{EveMir08}. Additionally, following the rationale behind random matrix theory (RMT), we assume that phases and norms of each wavefunctions in the correlator $C_{\alpha\beta}$ are independent random variables, so that only terms where phase factors cancel do survive the disorder average.\\

{\it Stationary contrast and $D_2$. }
The stationary contrast $\Lambda_\infty(x,x_0)$ defined in Eq.~\eqref{lambdaxx0inf} can be expanded in the momentum basis as $\frac{1}{N} \sum_\alpha \sum_{p_1p'_1p_2p'_2} C_{\alpha\alpha} \, e^{i[(p_1-p'_1)x+(p_2-p'_2)x_0]}-1$. Under the RMT assumption, the only non-vanishing terms left after disorder-average are those with $p_1=p'_1,p_2=p'_2$ and $p_1=p'_2,p_2=p'_1$. Taking care of double counting ($p_1=p'_2=p_2=p'_1$) and making use of normalization ($\sum_p |\varphi_\alpha(p)|^2=1$), we find
\begin{equation}
\label{lambdainf}
\Lambda_\infty(x,x_0) =\frac{1}{N} \sum_\alpha\sum_{p_1 \neq p_2}\big\langle \qty|\varphi_\alpha(p_1)|^2 \qty|\varphi_\alpha(p_2)|^2\big\rangle\, \e{i(p_1-p_2)(x-x_0)}.
\end{equation}
The contrast {\it at the tip of the peak}, $\Lambda_\infty \equiv \Lambda_\infty(x_0,x_0)$, can be evaluated by rewriting Eq.~\eqref{lambdainf} for $x=x_0$ as a sum over $p_1, p_2$ and subtracting its diagonal part. In contrast with systems with a mobility edge such as the Anderson model, all eigenvectors here have the same multifractal properties. The sum over $\alpha$, which is an average over eigenvectors, is then easily taken care of and we find:
\begin{equation}
\label{lambdainfx0}
\Lambda_\infty-1 = -\sum_p \langle|\varphi_\alpha(p)|^4\rangle \propto N^{-D_2},
\end{equation}
where $\varphi_\alpha(p)$ is an arbitrary eigenvector. The right-hand side of \eqref{lambdainfx0} is then obtained using
the well-known multifractal scaling of the inverse participation ratio (IPR) \cite{EveMir08}. Thus, remarkably, the CFS contrast is directly related to the multifractal dimension $D_2$. The prediction Eq.~\eqref{lambdainfx0} is very well verified in our model (see Fig.~\ref{fig:contraste-infini}). 
The contrast \textit{around the peak} can be obtained in the same manner from Eq.~\eqref{lambdainf} by using the multifractal scaling of the correlation function,  $\langle|\varphi_\alpha(p_1)|^2 \qty|\varphi_\alpha(p_2)|^2\rangle \sim |p_1-p_2|^{D_2-1}/N^{D_2+1}$ \cite{FyoMir97, CueKra07}, which yields
\begin{equation}
\label{lambdainfx}
\Lambda_\infty(x,x_0)\propto \frac{1}{N}\sum_{p=1}^{N-1}  \cos\qty[p(x_0-x)] \left(1-\frac{p}{N}\right) \qty(\frac{p}{N})^{D_2-1}.
\end{equation}
Here again, this general prediction directly links $\Lambda_\infty(x,x_0)$ to the multifractal dimension $D_2$. Note that Eq.~\eqref{lambdainfx} can actually be seen as a power-law decay $\sim |x-x_0|^{-D_2}$ for $x$ close to $x_0$, see Supp.~Mat.~\cite{SuppMat}.
As shown in Fig.~\ref{fig:contraste-infini}, Eq.~\eqref{lambdainfx} is also in very good agreement with numerical results for the RS model and reproduces quite well the spatial profile of the contrast in the region around $x=x_0$. 

Remarkably, the behavior Eq.~\eqref{lambdainfx0} can even be checked analytically in the RS model. Indeed, using a perturbation expansion at finite $N$ in the regime of strong multifractality $a\ll 1$, we get at first order in $a$ \cite{papierlong} the expressions $D_2=a$ and 
\begin{equation}
\label{finaloverlap}
\Lambda_\infty(x,x_0)= 2D_2 \sum_{p=1}^{N-1} \, \frac{\pi(1-\frac{p}{N})}{N\sin\frac{p\pi}{N}} \sin [p(x+\frac{\pi}{N})]\sin [p(x_0+\frac{\pi}{N})],
\end{equation}
which for $x=x_0$ leads to $\Lambda_\infty \sim D_2\log N$. Since $1-\gamma N^{-D_2}\sim  D_2\log N$ for $a\ll 1$, Eq.~\eqref{lambdainfx0} is verified analytically at first order in $a$ for the RS model. Note that the dip at $x=-x_0$ in Fig.~\ref{fig:contraste-infini}c, which is an idiosyncrasy of our model, is well-described by Eq.~\eqref{finaloverlap} (see \cite{papierlong} for details).

In the stationary limit $t \rightarrow \infty$ taken at finite system size $N$, the spatial profile of the CFS peak for a system with multifractal eigenstates is thus controlled by the multifractal dimension $D_2$. In particular, Eq.~\eqref{lambdainfx0} shows that the peak height value $\Lambda_\infty=1$ that was found in \cite{Lee14, Ghosh14, Ghosh15, Ghosh17} for disordered models and in \cite{Lema17} for the kicked rotor in the localized regime when $N\gg \xi \gg 1$, is reached here with an algebraic finite-size correction $N^{-D_2}$, a signature of multifractality. \\


\begin{figure*}[!t]
\centering
\includegraphics{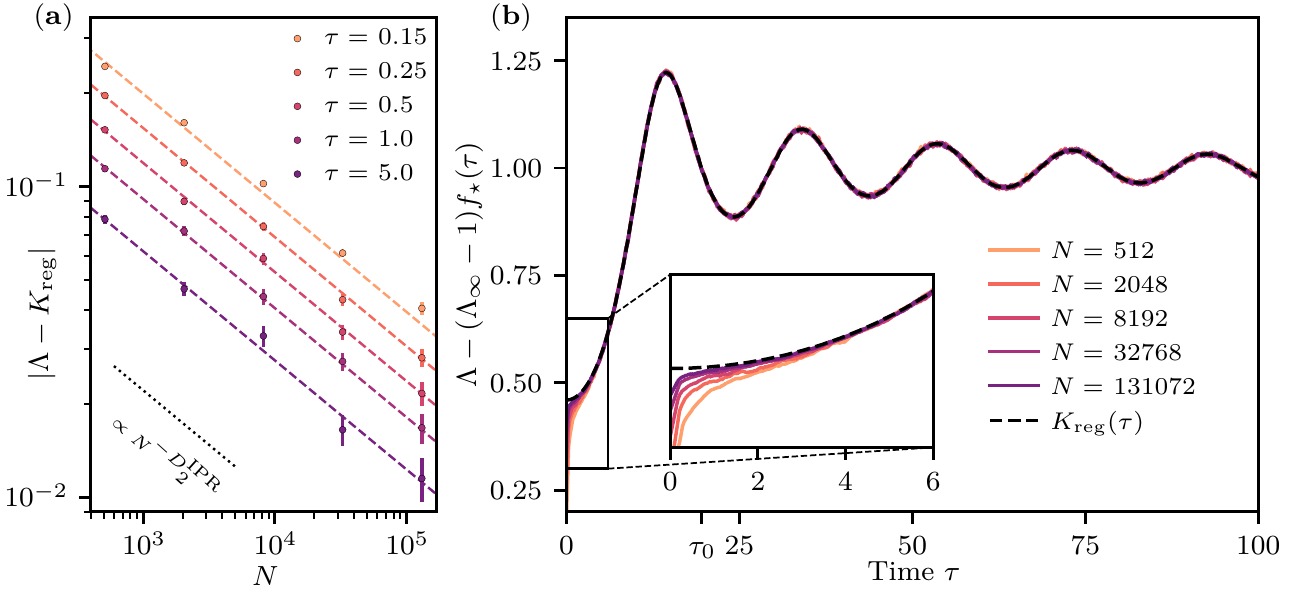}\includegraphics{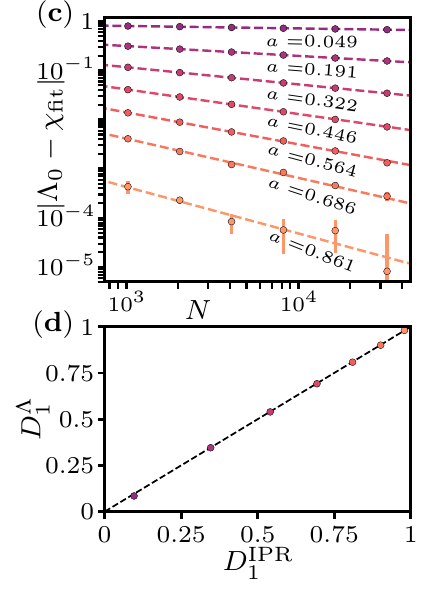}
\caption{Dynamics of the CFS peak at $x=x_0$. (a) Finite-size scaling of the difference between {the} numerically-computed contrast and $K_\text{reg}$, Eq.~\eqref{ffruij}, for $a=0.322$ and different $\tau=2\pi t/N$. Dashed lines correspond to {the} nonlinear fit $|\Lambda-K_\text{reg}|=\alpha N^{-D_2^\text{IPR}}$, where $D_2^\text{IPR}$ is independently determined from {a finite-size scaling of the IPR and $\alpha$ is the fitting parameter}. (b) Dynamics of the contrast for different $N$. Solid lines are numerical data smoothed over $\Delta \tau =0.3$. $C_0=0.55$ in $f_\star(\tau)$ (see text) is numerically extracted from $\hat{C}(\omega)$ (see Supp.~Mat.~\cite{SuppMat}). $\Lambda_\infty$ for large $N$ is extrapolated from Fig.~\ref{fig:contraste-infini}a. Black dashed line is $K_\text{reg}$ given by Eq.~\eqref{ffruij}. (c) Finite-size scaling of the difference between {the} contrast at $\tau=0.5$ and {the} level compressibility $\chi_\text{fit}$ for different {values of} $a$. $\chi_\text{fit}$ is extracted from a nonlinear fit $\Lambda=\chi_\text{fit}+\alpha N^{-D_2^\text{IPR}}$. (d) $D_1=1-\chi_\text{fit}$ {versus} $D_1$ {obtained from a } finite-size scaling of the wavefunction moments. When not visible, error bars are smaller than the symbol size.\label{contraste-taufini}}
\end{figure*}

{\it Time dynamics of the CFS peak height and $D_1$.}
We now aim at describing the temporal evolution of the CFS peak height $\Lambda (x_0,x_0,t)=\Lambda_\infty+F(x_0,x_0,t)$.
Starting from Eq.~\eqref{defb} and assuming eigenvector and eigenvalue decorrelation under disorder average when $x=x_0$, we get
\begin{equation}
\label{defbApprox}
\hat{F}(x_0,x_0,\omega) = \frac{2\pi}{N} \, \hat{R}(\omega) \, \displaystyle{\sum_{\alpha\neq\beta}}\big\langle\delta(\omega-\omega_{\alpha\beta}) \big\rangle , 
\end{equation} 
where the correlator $\hat{R}(\omega)=N^2 \big\langle |\varphi_\alpha(x_0)|^2 |\varphi_\beta(x_0)|^2\big\rangle_{\omega_{\alpha\beta}=\omega}$ only involves eigenfunctions whose quasi-energies are exactly separated by $\omega$ and does not depend on the labels $\alpha$ and $\beta$ because of disorder averaging. 
This implies that we can write the contrast as a convolution product
\begin{equation}
\label{CFSdyn}
\Lambda (x_0,x_0,t)=\Lambda_\infty+ [(K_N-1) \otimes R](t),
\end{equation}
where $R(t)$ is the inverse Fourier transform of $\hat{R}(\omega)$ and $K_N(t)=\langle \frac{1}{N}\vert\tr U^{t}\vert^2\rangle = 1 + \frac{1}{N} \langle \sum_{\alpha \neq \beta}e^{i\omega_{\alpha\beta} t}\rangle$ is the spectral form factor. 

To compute $R(t)$ in Eq.~\eqref{CFSdyn}, we follow the same steps as in the previous section : We expand the correlator $\hat{R}(\omega) = \sum_{p_1p'_1p_2p'_2} C_{\alpha\beta} (p_1,p'_1,p_2,p'_2) \, e^{i[(p_1-p'_1+p_2-p'_2)x_0]}$ in momentum space, where $C_{\alpha\beta}$ in Eq.~\eqref{lambdaxx0infexp} is computed for eigenfunctions with $\omega_{\alpha\beta}=\omega$, and only keep terms surviving disorder average. We find $\hat{R}(\omega)=1-\sum_p\expval{|\varphi_\alpha(p)|^2 |\varphi_\beta(p)|^2}_{\omega_{\alpha\beta}=\omega}$.
Writing $\sum_p\expval{|\varphi_\alpha(p)|^2|\varphi_\beta(p)|^2}_{\omega_{\alpha\beta}=\omega}=\sum_p\expval{|\varphi_\alpha(p)|^4} \hat{C}(\omega)$, three regimes can be identified for multifractal wavefunctions \cite{FyoMir97,EveMir08}:
\begin{equation}
\label{correlBs}
\hat{C}(\omega)= C_0\begin{cases}
1 &\omega < \omega_0\\
\qty(\omega/\omega_0)^{D_2-1}  &\omega_0 \leq \omega \leq \omega_1
\\
N^{D_2-1}\qty(\omega/\omega_1)^{-2} & \omega_1 \leq \omega,
 \end{cases}
\end{equation}
where $\omega_0$ is proportional to the mean level spacing $2\pi/N$, $\omega_1 \propto N\omega_0$, and $C_0$ is some numerical factor. We checked numerically that such an $\omega$-dependence is well-verified in the RS model with $\omega_0 = 2\pi a /N$ and the caveat that only the last two regimes are visible (see \cite{SuppMat,papierlong}) since there are no eigenstates separated by $\omega < \omega_0$ for this model \cite{OG030}.

We now note that the form factor at large $N$ is well approximated by the continuum limit $K_N(t) = \delta(\tau)+K_{reg}(\tau)$ \cite{Haa91} where the underlying $N$ dependence only appears through the scaled time $\tau = 2\pi t/N$. For the RS model, $K_{reg}(\tau)$ can be obtained analytically \cite{OG030} and reads:
\begin{equation}
\label{ffruij}
K_{reg}(\tau) = \frac{(1-a)^2 (a\tau)^2}{a^2(1-\cos a\tau)^2 + (a \sin a\tau +(1-a) a\tau )^2}.
\end{equation}
The sinusoidal terms in Eq.~\eqref{ffruij} come from the existence of a nonzero minimal level spacing in the RS model and are actually responsible for the temporal oscillations of the contrast. In the following, we will assume that this continuum limit holds for the form factor.


Let us now discuss the main result of this section. We rewrite Eq.~\eqref{CFSdyn} as 
\begin{equation}
\Lambda(x_0,x_0,\tau)-K_\text{reg}(\tau)=\delta(\tau)+  (\Lambda_\infty-1) f(\tau),
\end{equation}
where 
\begin{equation}
f(\tau)=1+C(\tau)+[(K_\text{reg}-1) \otimes C](\tau).
\end{equation}
Here $C(t)$ is the inverse Fourier transform of $\hat{C}(\omega)${. From Eq.~\eqref{correlBs}, we see that $C(\tau)$ and $f(\tau)$ do not depend on $N$ for $\tau>\tau_1\propto \frac{1}{N}\to 0$}.
After a very short fixed time $t_1=N\tau_1/2\pi$ ($\sim 1/a$ for the RS model), we thus get
\begin{gather}
\label{lambdatempx0}
    \Lambda(x_0,x_0,\tau)-K_\text{reg}(\tau) \propto N^{-D_2}.
\end{gather}
This scaling law can be seen as a generalization of Eq.~\eqref{lambdainfx0} at any time, and is very well verified in our model (see Fig.~\ref{contraste-taufini}a). 
At $\tau \gg \tau_0=2\pi/(N\omega_0)$, because of the plateau in Eq.~\eqref{correlBs}, we can even get an explicit approximation for $f(\tau)$ as $f_\star(\tau)=1+C_0(K_\text{reg}(\tau)-1)$; this gives $\Lambda(x_0,x_0,\tau)-(\Lambda_\infty-1)f_\star(\tau)=K_\text{reg}(\tau)$, 
as illustrated in Fig.~\ref{contraste-taufini}b,
which is a further illustration that $\Lambda(x_0,x_0,\tau)$ goes to $K_\text{reg}(\tau)$ at large $N$.

Noticeably, Eq.~\eqref{lambdatempx0} actually implies that in the {\it thermodynamic limit} ($N,t \to \infty$ at fixed $\tau$) the CFS contrast $\Lambda(x_0,x_0,\tau)$ is simply given by $K_{reg}(\tau)$ (which is the same result as in the localized regime \cite{Lee14, Ghosh14, Ghosh17}, because multifractal {finite-size effects vanish} in this limit).
In particular, at $\tau\to 0^+$, the CFS contrast converges to the level compressibility $\chi =K_{reg}(\tau\to 0^+)$ when increasing the system size $N$ (see Fig.~\ref{contraste-taufini}c).
We recover here, and actually demonstrate in a very general framework, the relation that was recently conjectured at the {Anderson transition} \cite{Ghosh17}, with an infinite system size at any fixed time $t$ ({implying} $\tau\to 0^+$).
The link between the level compressibility and the multifractal dimensions $D_q$ has a long {and} controversial history \cite{ChaKraLer96,EveMir08}. {However, the simple identity $\chi=1-D_1/d$ proposed in \cite{BogGirConj}, with $d$ the dimension of the system, was verified both analytically and numerically in various systems \cite{OG032,OG034}}, in particular in the RS model. 
Assuming this identity, together with the above considerations demonstrate that the CFS contrast at any fixed non zero time (or equivalently for $\tau \rightarrow 0^+$) goes to $\chi$ for $N\rightarrow \infty$ and thus gives direct access to $D_1$, as numerically demonstrated in Fig.~\ref{contraste-taufini}d:
\begin{equation}
\label{identiteD1}
\lim_{N \rightarrow \infty}\Lambda (x_0,x_0,t) =  1-D_1.
\end{equation}

{\it Conclusion.} In this Letter, we have shown that the CFS peak, a distinctive signature of Anderson localization, is also a marker of quantum multifractality, a fundamental property of several systems, such as the Anderson model at the metal-insulator transition. Our results are obtained for Floquet systems, but we believe them to be generic for critical disordered systems. Our work represents another situation which highlights the importance of studying physical observables in the space reciprocal to the space of localization/multifractality \cite{Vol18,Cher15,Sco20}.
Our results show that the multifractal dimensions $D_1$ and $D_2$ can unambiguously be measured from different dynamical quantities related to the CFS contrast. More precisely
(i) the stationary shape of the CFS peak gives access to $D_2$ via Eq.~\eqref{lambdainfx},
(ii) the finite-size scaling of the CFS contrast at any time gives $D_2$, Eqs.~\eqref{lambdainfx0} and \eqref{lambdatempx0}, a feature which had not been investigated previously, and
(iii) the limiting value of the CFS contrast at small $\tau$ gives $D_1$, Eq.~\eqref{identiteD1}.
Remarkably, the CFS contrast is actually a direct state-of-the-art experimental observable, as was recently observed in the localized regime with a kicked-like system, using standard time-of-flight techniques \cite{Hai18}. In such kicked-like systems, the system size $N$ can be precisely controlled, which should make it possible to reveal and monitor finite-size effects. 
Although the model \eqref{ruij} is difficult to implement directly with cold atoms because of the discontinuity of the potential \cite{pertur1,pertur2}, we think a suitably chosen temporal modulation could reproduce its main properties, as in  \cite{Hai18}. In addition, it could also be implemented with photonic crystals \cite{photonics1,photonics2,photonics3}.
Furthermore, we believe our results are generic enough to be relevant to other experimental systems where quantum multifractality is predicted to appear. 
As is well known, quantum simulations experiments are often plagued by dephasing mechanisms that destroy quantum coherence at long times (note that in \cite{Akridas2019} it was shown that wavepacket dynamics in momentum space only reveal multifractal properties at very long times). Our results suggest that this complication can be circumvented in our case since measurements can be done at short times. Our study thus paves the way to direct and robust measurements of multifractal properties of a quantum system that are notoriously hard to access by other means. Since multifractality is also known to appear in interacting systems~\cite{amini2014, Burmistrov2013,atas1,atas2}, a possible extension of this work could be to study the fate of CFS in the presence of interactions and their impact on our results.

\section*{Acknowledgments.}
C.M. wishes to thank S. Cristofari for stimulating discussions and Laboratoire Collisions Agr\'egats R\'eactivit\'e and Laboratoire de Physique Th\'eorique (IRSAMC, Toulouse) for their kind hospitality. This study has been supported through the EUR grant NanoX No.~ANR-17-EURE-0009 in the framework of the "Programme des Investissements d'Avenir", and research funding Grants No.~ANR-17-CE30-0024, ANR-18-CE30-0017 and ANR-19-CE30-0013. We thank Calcul en Midi-Pyrénées (CALMIP) for computational resources and assistance.


\end{document}